\documentclass{aa} 

\usepackage{graphicx}
\usepackage{txfonts}
\usepackage[colorlinks=true, citecolor=blue]{hyperref}
\usepackage[english]{babel}
\usepackage{mathtools} 
\DeclareMathOperator*{\argmin}{argmin}
\DeclareMathOperator*{\sgn}{sgn}
\DeclareMathOperator*{\card}{card}
\DeclareMathOperator*{\rank}{rank}
\usepackage{gensymb}

\begin{document}

    \title{Low-rank plus sparse decomposition for exoplanet detection 
    in direct-imaging ADI sequences}
    \subtitle{The LLSG algorithm}

        \author{C. A. Gomez Gonzalez
                \inst{1}
         \and{O. Absil}
         \inst{1}\fnmsep\thanks{F.R.S.-FNRS Research Associate}
         \and{P.-A. Absil}
         \inst{2} 
         \and{M. Van Droogenbroeck}
         \inst{3}
         \and{D. Mawet}
         \inst{4}
         \and{J. Surdej}
         \inst{1}\fnmsep\thanks{Honorary F.R.S.-FNRS Research Director}
                }

    \institute{Institut d'Astrophysique et de Géophysique, Université de Liège, Allée du Six Août 19c, B-4000 Liège, Belgium
                \and{Department of Mathematical Engineering, Universit\'e Catholique de Louvain, B-1348 Louvain-la-Neuve, Belgium}
                \and{Montefiore Institute, Université de Liège, B-4000 Liège, Belgium}
         \and{Department of Astronomy, California Institute of Technology, Pasadena, CA 91125, USA}
                }

    \date{Received September 19, 2015; accepted January 25, 2016}

 
   \abstract  
    {Data processing constitutes a critical component of high-contrast exoplanet 
    imaging. Its role is almost as important as the choice of a coronagraph or a 
    wavefront control system, and it is intertwined with the chosen observing strategy. 
    Among the data processing techniques for angular differential imaging (ADI), 
    the most recent is the family of principal component analysis (PCA) based 
    algorithms. It is a widely 
    used statistical tool developed during the first half of the past century. 
    PCA serves, in this case, as a subspace projection technique for constructing 
    a reference point spread function (PSF) that can be subtracted from the 
    science data for boosting the detectability of potential companions present 
    in the data. 
    Unfortunately, when building this reference PSF from the science data 
    itself, PCA comes with certain limitations such as the sensitivity of the lower 
    dimensional orthogonal subspace to non-Gaussian noise.}
    {Inspired by recent advances in machine learning algorithms such as robust
    PCA, we aim to propose a localized subspace projection technique that surpasses 
    current PCA-based post-processing algorithms in terms of the detectability of 
    companions at near real-time speed, a quality that will be useful for future 
    direct imaging surveys.}
    {We used randomized low-rank approximation methods recently 
    proposed in the machine learning literature, coupled with entry-wise 
    thresholding to decompose an ADI image sequence locally into low-rank, 
    sparse, and Gaussian noise components (LLSG). This local three-term decomposition
    separates the starlight and the associated speckle noise from the planetary
    signal, which mostly remains in the sparse term. We tested the performance of our new 
    algorithm on a long ADI sequence obtained on $\beta$ Pictoris with VLT/NACO.} 
    {Compared to a standard PCA approach, LLSG decomposition reaches 
    a higher signal-to-noise ratio and has an overall better performance in 
    the receiver operating characteristic space.
    This three-term decomposition brings a detectability boost compared to the
    full-frame standard PCA approach, especially in the small inner working angle 
    region where complex speckle noise prevents PCA from discerning true companions 
    from noise. }
    {}  

    \keywords{Methods: data analysis - 
                   Techniques: high angular resolution - 
                   Techniques: image processing - 
                   Planetary systems - Planets and satellites: detection}
                        
    \maketitle

\section{Introduction}  

	Only a small fraction (less than 2\%) of the confirmed exoplanet candidates
	known to date have been discovered through direct imaging 
	\footnote{http://www.exoplanet.eu}. 
	Indeed the task of finding exoplanets around their host stars with direct 
	observations is very challenging. The main difficulties in direct imaging from
	the ground are the image degradation caused by the Earth's turbulent 
	atmosphere, the huge difference in contrast between the host star and its 
	potential companions (typically ranging from $10^{-3}$ to $10^{-10}$), and 
	the small angular separation between them. 
	Detecting close-in planets with high-contrast imaging nowadays relies on four 
	main pillars \citep{mawet12}:  
	\begin{enumerate}
		\item optimized wavefront control,
		\item state-of-the-art coronagraphs,
		\item appropriate observing strategies,
		\item dedicated post-processing algorithms.
	\end{enumerate}
    
        The role of the coronagraphs is to suppress starlight and reduce the contrast
        between the star and its potential companions, while wavefront control helps
        reduce the amount and variability of diffracted starlight by correcting the 
        distorted wavefront in real time. 
        Unfortunately, even when combining these two technologies, the images suffer 
        from quasi-static speckle noise originating in the telescope and 
        instruments \citep{marois03spec}. These speckles, produced by slow variations in 
        the instrument aberrations, vary on timescales of several minutes to several 
        hours and follow modified Rician statistics \citep{soummer04,fitzg06} instead of 
        having a Gaussian nature. Moreover, these quasi-static speckles mimic the presence 
        of point-like sources (since they are comparable in angular size and brightness)
        and dominate the close vicinity of the star. Therefore, the need for optimal 
        observing strategies and fine-tuned post-processing procedures becomes evident.
        
        Observing strategies such as angular differential imaging \citep{marois06adi} 
        and spectral differential imaging using multiple spectral channels 
        \citep[SDI,][]{sparks02} allow decoupling, on the image plane, the planetary 
        signal from the speckle noise field, a situation that can be exploited by 
        algorithms of different complexities. In ADI the images are acquired with an 
        altitude/azimuth telescope in pupil-tracking mode, which means that the instrument 
        field derotator remains off, thereby keeping the instrument and telescope optics 
        aligned while the image rotates with time. For each image, a reference point-spread 
        function (PSF) can be constructed from appropriately selected images contained in 
        the same ADI sequence in such a way that, when it is subtracted from the image, 
        it reduces the stellar contribution and the associated quasi-static speckle noise. 
        Another consequence is that owing to the field-of-view rotation, the residual noise 
        averages incoherently after rotating the images to a common north. From the central 
        limit theorem, the      noise in the final combined image
mostly becomes 
        Gaussian \citep{marois08cl,mawet14ss}.
        
        Different approaches exist for generating this reference PSF from ADI sequences, 
        taking     the basic idea into account that the planet moves in circular 
        trajectories with respect to the speckle field. This is where post-processing 
        algorithms come into play as a critical step for boosting the detectability of 
        faint true companions in noisy background.      
        In ADI, as originally conceived, the reference PSF is constructed  
        by taking the median of the reference images. Some improvement can be 
        achieved by processing the frames in an annulus-wise fashion and applying a 
        rotation criterion for selecting the references. 
        LOCI \citep[locally optimized combination of images, ][]{lafren07} and
        its various modifications \citep{marois14, wahhaj15} aim to create a 
        reference PSF as a linear combination of the reference images independently 
        inside multiple subregions, in which the residual noise is minimized. 
        This approach provides a significant gain in sensitivity compared to the 
        classical median reference PSF. ANDROMEDA \citep[ANgular Differential OptiMal 
        Exoplanet Detection Algorithm,][]{mugnier09,canta15} treats the ADI sequence 
        in a pair-wise way and ensures that the images are chosen close enough in 
        time to guarantee the stability of the speckle noise and thereby allow its 
        suppression. In this approach the second image from every pair is used as a 
        reference PSF for the first.    
        
    More recently, principal component analysis (PCA) based algorithms 
    \citep{soummer12,amara12} have been proposed for reference PSF subtraction. The 
    reference PSF is constructed for each image as the projection of the image onto 
    a lower-dimensional orthogonal basis extracted from the references via PCA.
    The main advantages of this approach are that it can be applied to the 
    full images very efficiently using singular value decomposition (SVD), the reduced 
    number of free parameters (basically the size of the basis) and that PCA enables 
    forward modeling of astrophysical sources by fitting an astrophysical 
    model directly to the reduced images without introducing degeneracies \citep{soummer12}.
    
    Plain PCA extracts a lower-dimensional basis that is optimal in the least-square 
    sense. More recently, alternatives to the least-square criterion have been 
    proposed in the field of computer vision to consider other objectives, 
    such as sparsity of the noise or robustness to outliers. In this paper, we 
    propose one implementation of these algorithms applied to ADI image sequences. 
    In our approach we decompose the images locally into 
    low-rank, sparse, and Gaussian noise components to enhance residual speckle noise 
    suppression and improve the detectability of point-like sources in the final 
    combined image.
    
    Throughout the paper we use upper-case letters to denote matrices, 
    $\rank(X)$ to denote the rank of a matrix $X,$ and $\card(X)$ to denote the
    cardinality ($l_{0}$-pseudo norm or number of non-zero elements) of $X$.
      

\section{Subspace projection models and low-rank plus sparse decompositions}
        
        The problem of matrix low-rank approximation has been studied extensively 
        in recent years in many different fields, such as natural language processing, 
        bioinformatics, and computer vision. In particular for image analysis, there are 
        multiple tasks that can be achieved using low-rank modeling, such as image 
        compression, denoising, restoration, alignment, face recognition, and 
        background subtraction (or foreground detection in video sequences)
        \citep{zhou14lrmod}. 
        The applicability of low-rank 
        approximations is guided by the fact that the latent structure of high-dimensional 
        data usually lies in a low-dimensional subspace. If we consider a sequence of 
        $n$ images and a matrix $M \in \mathbb{R}^{m\times{n}}$ whose columns are 
        vectorized versions of those images, the above statement can be expressed as 
        $M=L+E$, where $L$ has low rank and $E$ is a small perturbation matrix. An 
        estimate of $L$ is given by a best low-rank approximation of $M$ in the 
        least-square sense:
    \[
    \min _{ L }{ \left\| M-L \right\|  }_{F}^{2} ,\, \text{subject to}\, \rank(L)\le k,
    \]
    where ${\left\| X \right\|}_{F}^{2}=\sqrt {\sum _{ij}{ X_{ij}^{2} }}$ 
    denotes the Frobenius norm of a matrix $X,$ and $k$ is the rank of the low-rank
    approximation $L$. This can be solved analytically through SVD \citep{eckart_young,candes}:
    \[ 
    M=U\Sigma V^{ T}=\sum _{i=1}^{k}{ \sigma_{i}u_{i}v_{i}^{T}  },  
    \]    
        where the vectors $u_{i}$ and $v_{i}$ are the left and right singular vectors,
        and $\sigma_{i}$ the singular values of $M$. Choosing the first $k$ left singular
        vectors forms an orthonormal basis for the low-dimensional subspace that
        captures most of the variance of $M$. This procedure corresponds to PCA 
        \citep{hotelling}, as it is usually called in statistics. 
                
        In computer vision, for the task of segmentation of an image sequence into 
        background and foreground pixels, PCA was proposed by \citet{oliver99},
        who modeled the background pixels using an eigenspace model.
        Each image is approximated by its projection onto the first $k$ principal 
        components. They noted that, because they do not appear at the 
        same location in the $n$ sample images and are typically small, moving objects do not 
        make a significant contribution to the PCA model. The foreground pixels are 
        found by subtracting from each image its low-rank PCA approximation 
        and thresholding the pixel values in the residual images. 
                
        In astronomy PCA has proven to be effective for modeling time- and 
        position-dependent PSF variations of the Sloan Digital Sky Survey and later 
        for the Advanced Camera for Surveys on the Hubble Space Telescope 
        \citep[see][]{jee}.
        In the context of reference PSF subtraction for high-contrast imaging, a PCA
-based approach has been proposed independently by \citet{soummer12} and
        \citet{amara12}. The problem of modeling and subtracting a reference PSF 
        with the purpose of detecting a moving planet in an ADI image sequence has
        a lot in common with the segmentation of video sequences into background and
        foreground pixels (e.g., for video surveillance and detection of moving objects), 
        since the reference PSF and quasi-static speckles can be modeled using a low-rank 
        PCA approximation. The orthogonal basis formed by the first 
    principal components (PCs) is learnt from the ADI sequence itself, which
    adds complications to the low-rank approximation task because some part of 
    the foreground signal is absorbed in the background model. This relates to the 
    fact that PCA gives a suitable low-rank approximation only when the term $E$
    (foreground signal) is small and independent and identically distributed 
    Gaussian \citep[see section 1.1 in ][]{candes}. This is unfortunately not
    the case for moving planets in ADI images.
    
\subsection{Robust PCA}   
    
        In recent computer vision literature, several subspace projection algorithms
        exploiting the low-rank structure of video sequences have been proposed to 
        solve the weaknesses of the basic PCA model and provide more versatile 
        and robust background models \citep{bouwmans14rpca}. 
        The most notable is the family of robust PCA (RPCA) algorithms, which model 
        the data as the superposition of low-rank and sparse components, containing 
        the background and the foreground pixels, respectively. 
        One of the first approaches for solving this decomposition was proposed by
        \citet{candes}, with an algorithm called principal component pursuit (PCP). PCP
        aims at decomposing $M$ into low-rank plus sparse ($L+S$) matrices by solving the 
        following problem:
        \[ 
        \min_{L,S} {\left\| L \right\|_{ * }}+\lambda  \left\| S \right\|_{1}, \,\text{subject to} \, L+S=M, 
        \]
        where $L$ is low-rank, $S$ contains sparse signal of arbitrarily large
        magnitude, ${\left\| S \right\|}_{1}=\sum_{ij} {\left| S_{ij} \right|}$ 
        is the $l_{1}$-norm of $S,$ and
        ${ \left\| L \right\|  }_{*}$ denotes the nuclear norm of $L$ or sum of its 
    singular values. The nuclear norm and the $l_{1}$-norm are the convex relaxations
        of $\rank(L)$ and $\card(S)$ and provide the best computationally tractable 
        approximation to this problem. Under rather weak
        assumptions, this convex optimization recovers
 the low-rank and sparse components that separate the varying 
        background and the foreground outliers. Important limitations of this algorithm
        are its high computational cost and the assumption that the low-rank component
        is exactly low-rank, and the sparse component is exactly sparse, contrary to what 
        we find in real data, which is often corrupted by noise affecting a large part of 
        the entries of $M$ \citep{bouwmans14rpca}. In the ideal case, when applying such 
        decomposition to an ADI image sequence, the reference PSF would be captured by 
        the low-rank component and the small moving planets (realizations of the 
        instrumental PSF) by the sparse component. In real ADI coronagraphic images, the 
        reference PSF, composed of the stellar PSF and speckles, is never exactly 
        low-rank owing to the quasi-static component of the speckle noise.
        Therefore the exact decomposition into low-rank and sparse components does not
        exist, and the $S$ component recovered by PCP becomes polluted by residual noise 
        from the quasi-static speckles that will produce a final image resembling the 
        results of standard PCA.
        
\subsection{GoDec}

        Several modifications of PCP have been proposed to address its limitations with
        real data for the problem of foreground detection \citep[see][for a 
        complete review]{bouwmans14rpca}. Beyond PCP, there are different 
        approaches to RPCA via low-rank plus sparse matrix decomposition. 
        
        Among them, GoDec \citep[Go Decomposition,][]{zhou11godec} is a convenient approach,
        in terms of computational cost, to the decomposition of $M$. It proposes 
        a three-term decomposition (instead of the typical low-rank plus sparse one):
        \[
        M=L+S+G, \rank(L) \le k, \card(S) \le c,
        \] 
        where $G$ is a dense noise component, and $k$ and $c$ the constraints on the rank
        of $L$ and the cardinality of $S$. GoDec produces an approximated 
        decomposition of $M$, whose exact low-rank plus sparse decomposition does not
        exist because of additive noise $G$, restricting $\rank(L)$ and $\card(S)$ in order 
        to control model complexity. This three-term decomposition can be expressed as 
        the minimization of the decomposition error:
        
        \begin{equation} \label{godec1} 
    \min_{L,S} {\left\| M-L-S \right\|^{2}_{F}}, \, \text{subject to} \, \rank(L)\le k, \, \card(S)\le c.
        \end{equation}

        The optimization problem of Eq. \eqref{godec1} is tackled by alternatively
        solving the following two subproblems until convergence, when the decomposition 
        error reaches a small error bound (=10$^{-3}$):

        \begin{equation} \label{godec2} 
    \begin{dcases} 
        L_{ t }=\argmin_{\substack{\rank(L)\le k }}{ \left\| M-L-S_{t-1} \right\|_{F}^{2}} ;\\
        S_{ t }=\argmin_{\card(S)\le c }{ \left\| M-L_{t}-S \right\|_{F}^{2}} . 
    \end{dcases}    
        \end{equation}

        In Eq. \eqref{godec2}, $L_{t}$ can be updated via singular value hard
        thresholding of $M-S_{t-1}$ (via SVD in each iteration), and $S_{t}$ via 
        entry-wise hard thresholding of $M-L_{t}$. 
        It must be noted that singular value hard thresholding is equivalent
        to the truncation of the number of PCs in the PCA low-rank approximation. 
        
        A randomized and improved version of GoDec was proposed by the same authors
        with SSGoDec. In this approximated RPCA algorithm, the cardinality constraint 
        is modified by introducing an $l_{1}$ regularization, which induces 
        soft-thresholding when updating $S$ \citep{zhou13}. The soft-thresholding 
        operator $\mathcal{S}_{\gamma}$ with threshold $\gamma$ 
        applied to the elements of a matrix $X$ can be expressed as
        \[
        \mathcal{S}_{\gamma}X = \sgn(X_{ij}) \max\left(\left| X_{ij} \right| - \gamma,0 \right).
        \]

        The reduced computational cost of SSGoDec mostly comes from using, on each 
        iteration, the bilateral random projections \citep{zhou11brp} of $M$ 
        instead of singular value thresholding for its low-rank approximation. BRP is 
        a fast randomized low-rank approximation technique making use of $M$'s left 
        and right random projections,
        $Y_{1}=MA_{1}$ and $Y_{2}=M^{T}A_{2}$, where $A_{1} \in \mathbb{R}^{n\times{k}}$ and
        $A_{2} \in \mathbb{R}^{m\times{k}}$ are random matrices. The rank-$k$ 
        approximation of $M$ is computed as
        \[ L=Y_{1}(A_{2}^{T}Y_{1})^{-1}Y_{2}^{T}. \] 
        The computation of this approximated $L$ requires less floating-point 
        operations than the SVD-based approximation. The bounds of the approximation
        error in BRP are close to the error of the SVD approximation under mild
        conditions \citep{zhou11brp}.


\section{Local randomized low-rank plus sparse decomposition of ADI datasets}
        
    \begin{figure*}[!t]
    \begin{center}
    \includegraphics[width=16cm]{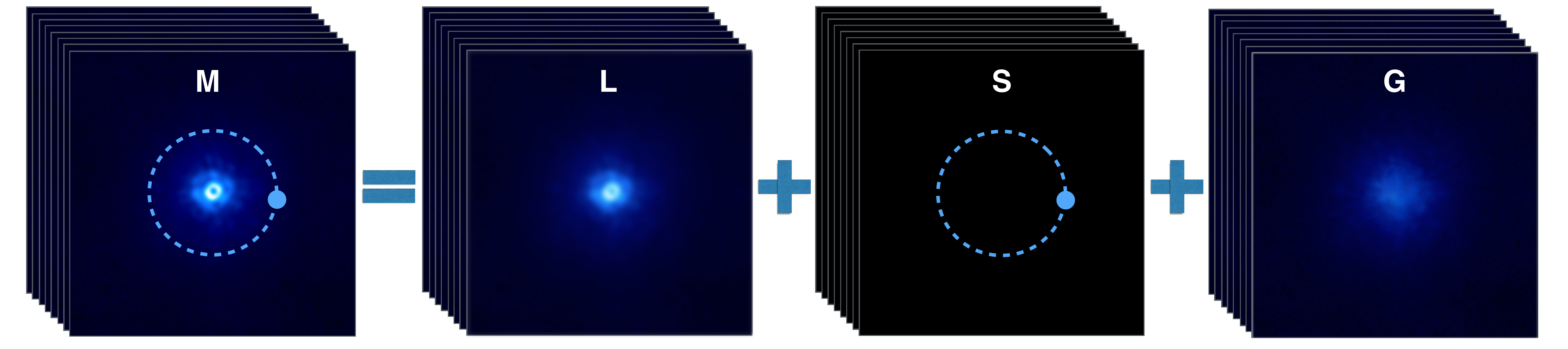}
       \caption{LLSG decomposition of ADI data (left) into low-rank (middle left) 
                        plus sparse (middle right) plus Gaussian noise (right) terms. In the ideal 
                        case, this decomposition separates the reference PSF and quasi-static 
                        speckle field from the signal of the moving planets, which stays in the 
                        sparse component.
               }
          \label{fig0}
    \end{center}
    \end{figure*}       
        
    \begin{figure}[!t]
    \begin{center}
    \includegraphics[width=5cm]{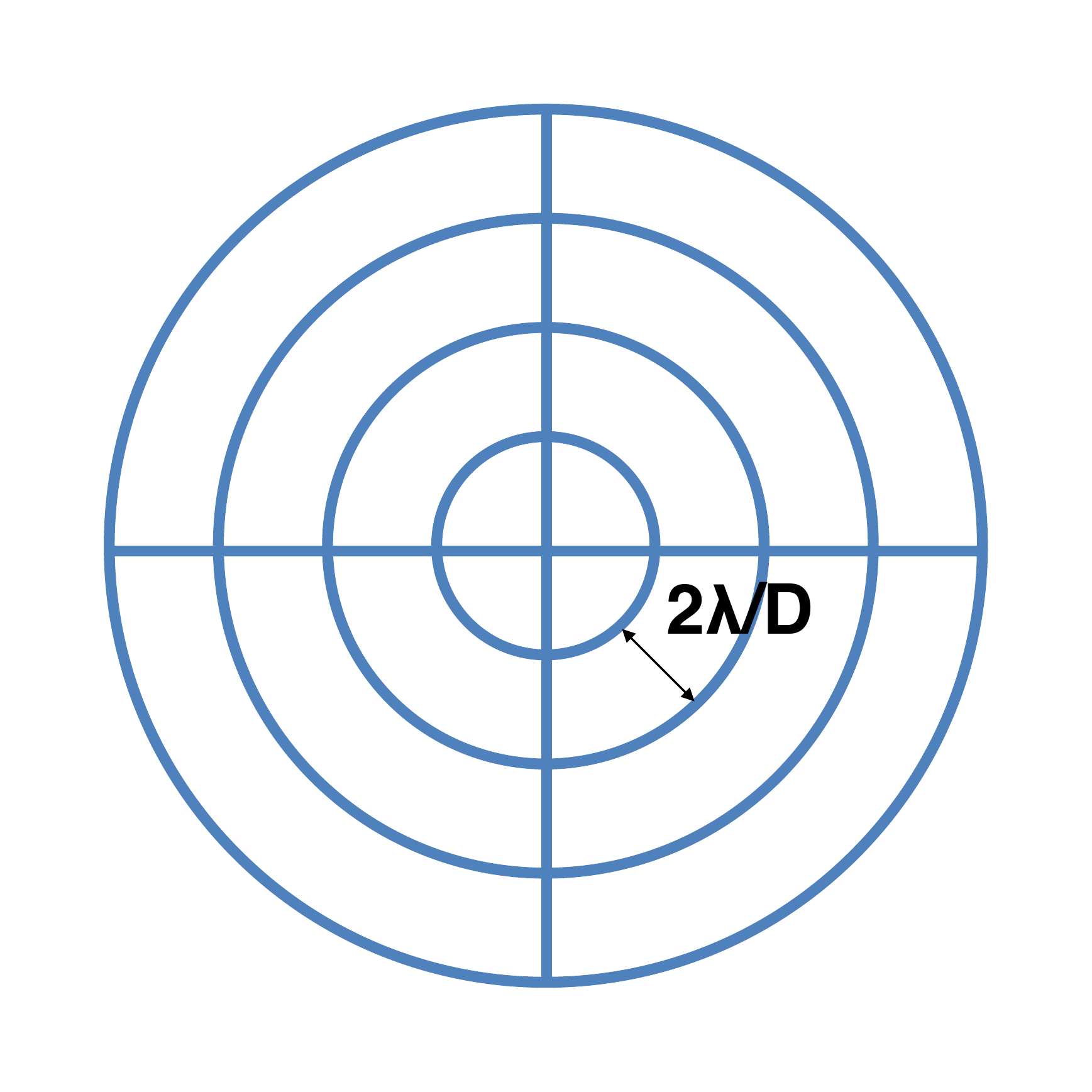}
       \caption{Quadrants of annuli used for partitioning the images.}
          \label{fig7}
    \end{center}
    \end{figure}                
        
        Restricting the cardinality of $M-L_{t}$ while operating on whole images is
        problematic in the presence of multiple companions, as the dimmest one could
        get severely subtracted from the data (especially for close-in companions), or 
        bright speckles could turn into false positives. We find that applying a local
        three-term decomposition, which exploits the geometrical structure of ADI 
        image sequences, can alleviate the problem of a global thresholding and in addition 
        provide a better low-rank approximation for the given patch.
        
        These ideas were put together to build an ADI post-processing algorithm
        for boosting point-like source detection, the Local Low-rank plus Sparse plus 
        Gaussian-noise decomposition (LLSG). A schematic illustration of this decomposition
        is shown in Fig. \ref{fig0}. The algorithm follows four main steps: 
        \begin{enumerate}
                \item the images of the cube are broken into patches, specifically in 
                quadrants of annuli of width 2$\lambda/D$ (see Fig. \ref{fig7});
                \item each of these quadrants is decomposed separately as in Eq. \eqref{godec2}, 
                alternatively updating its $L$ and $S$ components for a fixed number of iterations;
                \item for each patch, the $S$ component of the decomposition is kept;
                \item all frames are rotated to a common north and are median combined
                in a final image.
        \end{enumerate}                 
        The soft-thresholding will enforce the sparsity of the $S$ component 
        throughout the iterations. Instead of using a common threshold parameter $\gamma$, 
        we use a different one for each patch. 
         Values of $\gamma$ that are too high will remove the signal of companions too much along 
        with the residual speckle noise, while values that are too low will not lead to much 
        improvement over PCA processing, therefore hindering the detection of potential 
        very faint companions.
        Instead of leaving this as a free parameter, our algorithm defines $\gamma$ for 
        each patch as the square root of the median absolute deviation of the entries in 
        $S_{t}$. This thresholding can be scaled up or down safely by the user with a 
        tunable parameter in case it is needed. 
        Partitioning the frames using quadrants of annuli does not increase the computational
        cost and alleviates the problem of applying entry-wise soft thresholding 
        globally (on the whole frames), thereby giving better results in the case of several 
        companions with different brightnesses or when very bright speckles are present.

        Among the free parameters of LLSG, the rank (low integer value) is 
        certainly the most important one. This parameter is equivalent to the number of PCs 
        in the PCA algorithm and defines the size of the low-rank approximation of our dataset 
        ($L$ term). Values of the rank that are too high cause too much planetary signal 
        to be absorbed by the low-rank term, 
        whereas a low value produces a noisier sparse term. The sweet spot depends on data. 
        The sparsity level (for scaling the soft-thresholding, by default is equal to one) is 
        the second parameter of LLSG, which controls how sparse the $S$ term is
        and how much noise goes into the $G$ term.
        It usually does not require user intervention since it is internally defined 
        for each image patch. The third parameter of LLSG is the number of iterations. 
        A small number of iterations is enough (the default is ten) to achieve good 
        decomposition according to our tests on several datasets, but it can be fine-tuned 
        by the user. The number of iterations affects the running time of LLSG and 
        generally by doubling the number of iterations we double the computation time.   
        
        As explained before, the sparse component is the main product of this algorithm, 
        where potential companions will have a higher signal-to-noise ratio (S/N). The two other components in this
        decomposition, the low-rank and the noise, can serve as estimates of the
        total noise of our data. The $L$ component will contain the starlight and 
        most of the static and quasi-static structures, while $G$ will 
        capture the small and dense residual noise that was not captured by the 
        low-rank approximation. An implementation of LLSG for ADI data is 
        provided in the Vortex Image Processing (VIP) 
        pipeline\footnote{\href{https://github.com/vortex-exoplanet/VIP}{https://github.com/vortex-exoplanet/VIP}}. VIP is written in Python 2.7 programming language (Gomez Gonzalez 
        et al. in prep.)
        
        This idea of a three-term decomposition with some modifications, e.g. 
        a different partitioning of the images, could be used as well for spectrally 
        dispersed data obtained with an integral field spectrograph (IFS). 
        After rescaling IFS data, the companions will appear to move radially 
        through the speckle noise field. Therefore, LLSG can be a good choice for 
        decomposing the image sequence and capturing potential planets in the sparse term.
        

\section{Application to real data}

    \begin{figure*}[!t]
    \begin{center}
    \includegraphics[width=18cm]{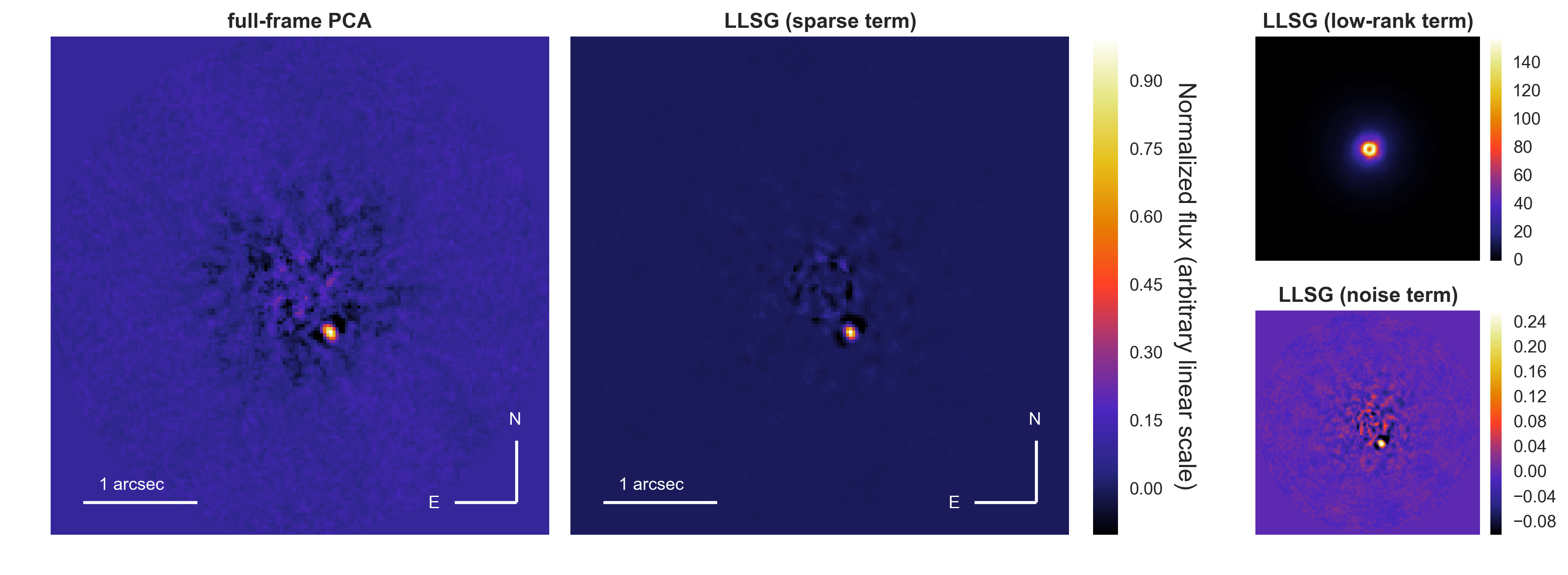}
       \caption{Final result of post-processing with PCA (left) and the three
                        terms of the LLSG decomposition (middle and right) for $\beta$ Pic 
                        NACO data. 38 PCs were used to maximize the S/N for $\beta$ 
                        Pic $b$ in the full-frame PCA approach. For the LLSG decomposition a 
                        rank of 10 gave a significant improvement on $b$'s S/N (see Fig. \ref{fig2}). 
                        All the frames were 
                        normalized to the maximum value of the LLSG sparse frame (middle panel).
               }
          \label{fig1}
    \end{center}
    \end{figure*}       
                
    \begin{figure*}[!t]
    \begin{center}
    \includegraphics[width=15cm]{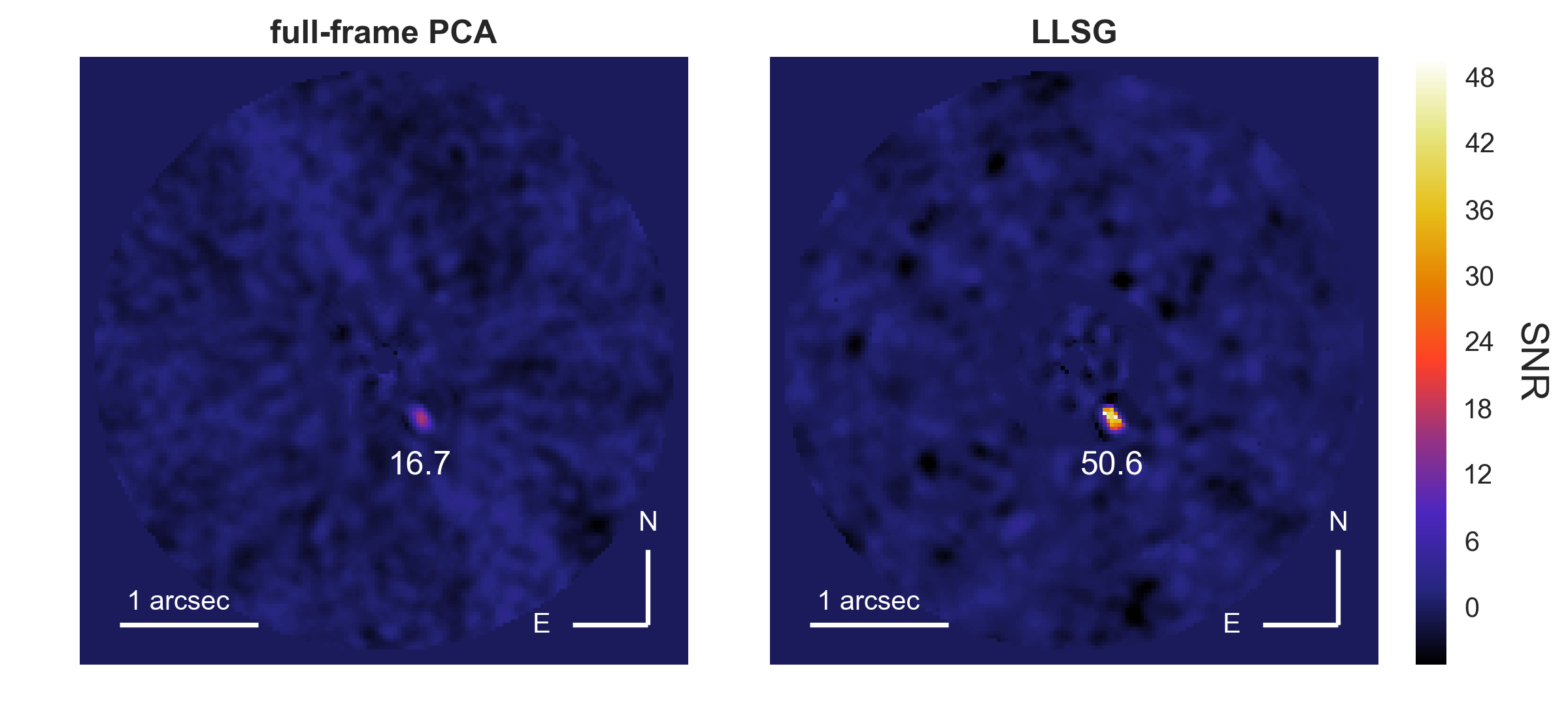}
       \caption{S/N maps for PCA (left) and LLSG (right) computed from the final frames
                        shown in Fig. \ref{fig1} (left and middle panels). Planet $\beta$ Pic $b$ 
                        has roughly three times higher S/N when processed with our LLSG decomposition.
               }
          \label{fig2}
    \end{center}
    \end{figure*}

\subsection{Data used}

        The application of the LLSG decomposition to real data gives a first taste of
        its capabilities. In this paper, we use the data set of $\beta$ Pic and its
        planetary companion $\beta$ Pic $b$ \citep{lagrange10} obtained on 
        January 2013 with VLT/NACO in its AGPM coronagraphic mode \citep{oabsil13}. 
        The observations made at $L'$ band were performed under poor conditions, 
        nevertheless $\beta$ Pic $b$ could be seen on the real-time display thanks to 
        the excellent peak starlight extinction provided by the AGPM. The total on-source 
        integration time was 114 min with a parallactic angle ranging from 
        $-$15$\degree$ to 68$\degree$. A clean cube was obtained after basic 
        preprocessing steps, such as flat fielding, bad pixel removal, bad frames removal, 
        recentering of frames, and sky subtraction. After temporal subsampling, by averaging 
        40 successive frames, a new cube of 612 individual frames with 8 sec of effective 
        integration time was created \citep[for details see][]{oabsil13}. As a final 
        step the central 161x161 pixels were cropped on each frame.

\subsection{Results}

        Figure \ref{fig1} shows the final post-processed frames using full-frame PCA and 
        the three terms of the LLSG decomposition. We see clearly how LLSG can separate 
        the starlight and quasi-static speckles from the planetary signal. The sparse 
        term is where most of the signal of $\beta$ Pic $b$ is present.
        In the following discussion, we use the S/N between the planet signal 
        and the background pixels to compare the performance of the two algorithms for 
        the task of detection of point-like sources.
        
        For calculating the S/N, we depart from the previously used definition 
        in high-contrast imaging where the pixels are assumed to be statistically independent
        and the S/N is basically considered as the ratio of the flux in an aperture
        centered on the planet to the standard deviation of the pixels in an annulus 
        at the same radius. We instead adopt the definition proposed by \citet{mawet14ss} 
        that considers the problem of small sample statistics applied to small-angle 
        high-contrast imaging. The number of resolution elements ($\lambda/D$) 
        decreases rapidly toward small angles, thereby dramatically affecting confidence 
        levels and false alarm probabilities. In this small sample regime, a two-sample 
        t-test is used to see whether the intensity of a given resolution element 
        is statistically different from the flux of similar $\lambda/D$ circular 
        apertures at the same radius $r$ from the star. 
        When one of the samples contains the single test resolution element, the two-sample t-test brings a 
        better definition of S/N and is given by
        \[
        S/N = \frac {\overline{x}_{1}-\overline{x}_{2}}{s_{2}\sqrt{1+\frac{1}{n_{2}}}}
        \]      
        where $\overline{x}_{1}$ is the intensity of the single test resolution 
        element, $n_{2}$  the number of background resolution elements at the 
        same radius ($n_{2} = round(2\pi r)-1$, where $r$ is given in $\lambda/D$ units) ,
        and $\overline{x}_{2}$ and $s_{2}$ are the mean intensity and the empirical 
        standard deviation computed over the $n_{2}$ resolution elements. 
        
        The use of the S/N as a metric for comparing algorithms can become problematic 
        when, in some cases, the noise can be almost totally suppressed, making the S/N 
        infinite. In this scenario, if a companion is present, a clear detection through visual 
        vetting can be claimed. We have encountered this situation when processing other 
        datasets of better quality (conditions of observation and/or better wavefront 
        sensing). We also note that, the S/N of a point-like source depends on the choice 
        of the aperture sizes and the position of the apertures themselves, especially at
        small angular separations where the small sample statistics effect becomes 
        dominant \citep{quanz15}. Throughout this paper we use an aperture 
        size of 4.6 pixels, which is the Gaussian full width at half maximum (FWHM) measured 
        on the off-axis PSF of $\beta$ Pic. 
 
        Also, the positioning of the apertures is done in an automatic way and is the same 
        for each realization, when measuring S/Ns on the final frames of the compared algorithms. 
        As an example, we only have 24 background apertures ($n_{2}$) for the case of
        $\beta$ Pic $b$ (using the FWHM as an approximation for the value of the $\lambda/D$
        parameter).
        In spite of these limitations, we stick to the use of the S/N for its 
        practicality for the task of detecting point-like sources. 
        
        Figure \ref{fig2} shows the S/N maps corresponding to full-frame PCA and the sparse 
        term of the LLSG algorithm. With LLSG the S/N of $\beta$ Pic $b$ is 
        roughly three times higher than with full-frame PCA thanks to the small amount of 
        residual noise in $S$. To maximize the S/N with full-frame PCA, we varied the number of 
        PCs in the interval from 1 to 100, measuring at every step the S/N at the location
        of $\beta$ Pic $b$. The highest S/N (=16.7) was achieved with 38 PCs. In 
        the case of LLSG, the best compromise between residual noise subtraction and 
        companion signal recovery was obtained with a rank equal to 10. The default number 
        of iterations worked well for this ADI sequence.        
        
        As we can see in Fig. \ref{fig1}, roughly 25$\%$ of the 
        planetary signal leaks into the LLSG noise term. However, this is less than the amount
        of companion signal absorbed in the PCA low-rank approximation, when using the 38
        PCs that maximized the S/N of $\beta$ Pic $b$. In this case, the leaking into the
        $G$ term does not hinder the goal of LLSG for improving the detectability of a 
        point-like source. In the following section, we test whether this holds true for more
        complicated scenarios with fainter companions. Nevertheless, the ultimate goal
        is to avoid any signal loss. This will be the subject of future work. 
                
\section{Simulations with synthetic companions}

    \begin{figure*}[!t]
    \begin{center}
    \includegraphics[width=18cm]{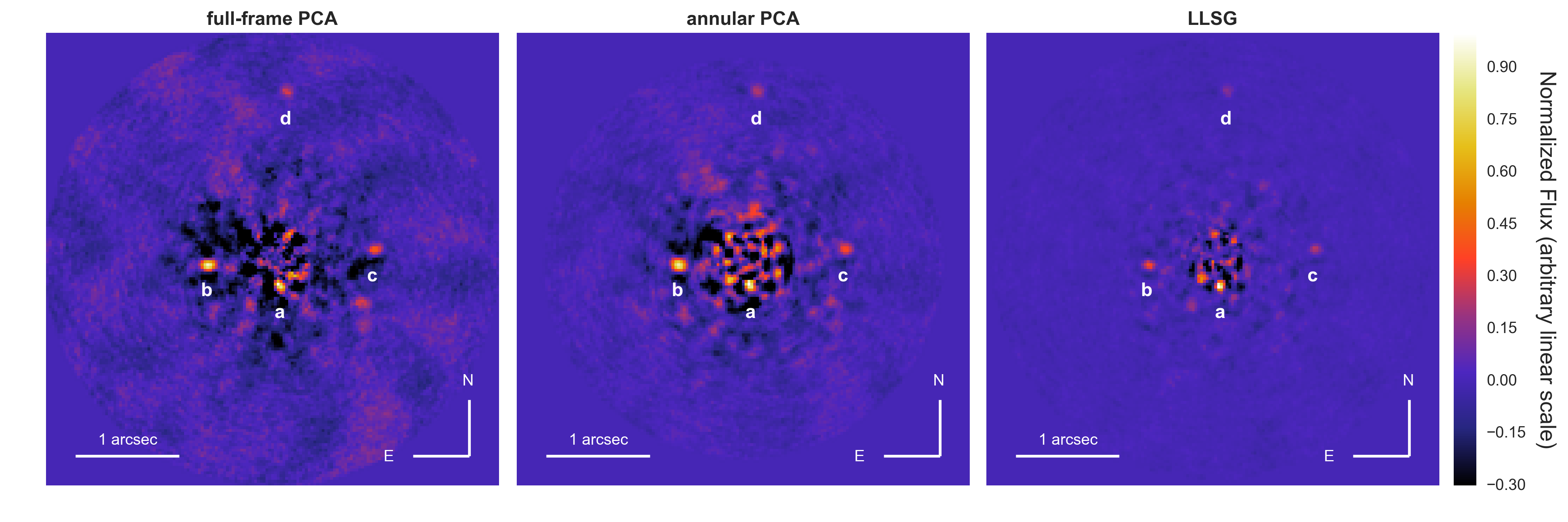}
       \caption{Final result of post-processing with PCA (left), annular PCA (middle),
                        and our LLSG decomposition (right). 
                        The images were normalized to their own maximum value. Four 
                        synthetic companions a(285$\degree$), b(185$\degree$), c(5$\degree$), 
                        and d(85$\degree$) were injected at 2, 5, 8, and 13 $\lambda/D$ from the 
                        center, respectively. The injected PSFs were scaled at 0.5, 5, 5, and 7 
                        times the noise of the respective annulus. 
                        We used 13 PCs when applying full-frame 
                        PCA and 25 PCs for the annular PCA (applying the same number of PCs in every
                        annulus) in order to maximize the S/N of the innermost fake companion in 
                        each case.}
          \label{fig3}
    \end{center}
    \end{figure*}       
                
    \begin{figure*}[!t]
    \begin{center}
    \includegraphics[width=18cm]{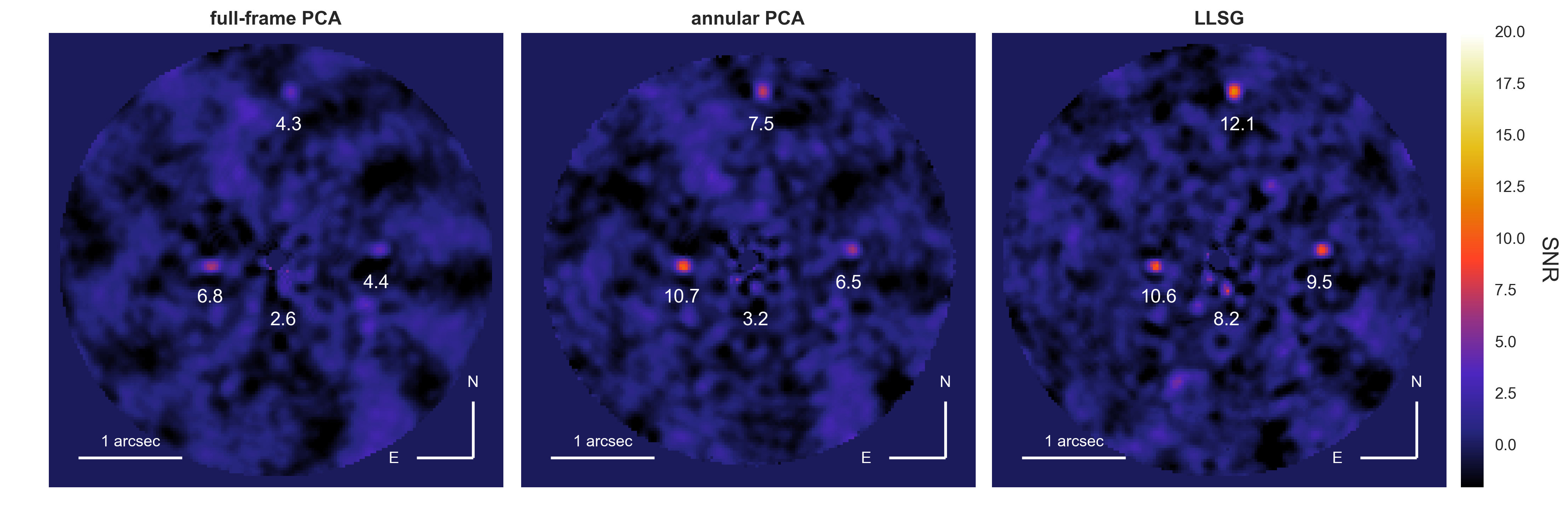}
       \caption{S/N maps for full-frame PCA (left), annular PCA (middle), and our 
                        LLSG decomposition (right) showing the values of each fake companion
                        S/N. With our algorithm the four injected
                        companions are clearly revealed. The S/N of the fake companion at 
                        2$\lambda/D$ is clearly at the level of the noise (false negative) 
                        in the case of full-frame PCA and its annular version. With LLSG we 
                        reach a peak S/N that is three times higher. For 
                        the rest of synthetic companions, the S/N obtained with LLSG 
                        is up to three times higher than the one obtained with full-frame PCA.}
          \label{fig4}
    \end{center}
    \end{figure*}

\subsection{Single test case}

        The use of on-sky data with simulated companions allows us to probe the 
        performance of the detection algorithms with planets at different locations
        on the image plane and with varying brightness.
        This enables us to test how LLSG deals with a fainter and closer-in 
        companion than $\beta$ Pic $b$, which presented a rather easy scenario.
        To obtain a data cube without any companion, $\beta$ Pic $b$ was 
        subtracted using the negative fake companion technique \citep{lagrange10, oabsil13},
        which uses the off-axis PSF as a template to remove the planet from each
        frame by optimizing the position and flux (of the injected negative candidate).
        This optimization is performed by minimizing the sum of the absolute values of the 
        pixels in a 4$\times$FWHM aperture on the PCA processed final frame. We used a 
        downhill simplex minimization algorithm \citep{nelder65} for this
        purpose, which is enough to obtain a planet-free cube.

        In the empty cube, we injected the normalized off-axis PSF to create four
        synthetic companions and compared the results of full-frame PCA and our approach 
        (see Figs. \ref{fig3} and \ref{fig4}). The companions a(285$\degree$), 
        b(185$\degree$), c(5$\degree$) ,and d(85$\degree$) were injected at 2, 
        5, 8, and 13 $\lambda/D$ from the center, respectively. The brightness of the fake
        companions was scaled as a function of the local noise before injection. 
        The noise was measured as the standard deviation of the fluxes of the resolution 
        elements inside the corresponding annulus in the classical ADI processed frame 
        (which means it has been median-subtracted, derotated, and median-combined). 
        The injected PSFs were scaled at 0.5, 5, 5, and 7 times the noise of 
        the respective annulus.
        
        In this particular example of processing a cube with several injected synthetic       
        companions, we encounter a first problem with full-frame PCA: it is not possible 
        to optimize the S/N for each individual companion at the same time by
        adjusting the number of PCs used. For the innermost injected planet, we 
        need to use 13 PCs for full-frame PCA in order to reduce the residual speckle 
        noise and achieve the best possible S/N. As done previously, the number 
        of PCs was varied from 1 to 40, each time measuring the S/N at the location where 
        the innermost planet was injected. This number of PCs may not be optimal for farther 
        companions, which could achieve higher S/Ns with a smaller number of PCs. The 
        optimal number 
        of PCs in general decreases when the planet is farther away from the star in 
        the photon-noise limited regime, since the planets have more rotation and the speckle 
        noise is not dominant.
        
        A better strategy in this case is to use the PCA low-rank approximation 
        annulus-wise (see middle panel in Fig. \ref{fig3}). In this case, 
        it is even possible to apply a frame-rejection criterion based on a parallactic 
        angle threshold \citep{oabsil13}. The motivation behind this is that an 
        annular PCA low-rank approximation will capture the background and speckle noise 
        in a better way for a given patch. Furthermore, 
        keeping only the frames where the planet has rotated by at least
        $1\lambda/D$ in our 
        PCA reference library, we prevent the planetary signal from being captured by the low-rank
        approximation and subsequently subtracted from the science images, thereby 
        increasing the S/N in the final frame. We provide a parallelized implementation 
        of this algorithm in the VIP repository. For the innermost planet, located at 
        $2\lambda/D$, we can obtain a maximum S/N of 3.2 after optimizing the number of 
        PCs (by testing from 15 to 35 PCs) and using $2\lambda/D$ wide annuli. For
        LLSG we kept the same rank as we used before and slightly reduced the sparsity level
        to achieve the highest S/N for the innermost fake companion.
        
        As seen in the S/N maps shown in Fig. \ref{fig4}, our LLSG algorithm provides a gain 
        of a factor three in S/N at $2\lambda/D$ with respect to full-frame PCA, resulting 
        in a clear detection instead of the false negative in the case of full-frame PCA 
        (even after careful optimization of the PCA truncation and knowledge of the 
        coordinates of the planet). For the three other synthetic companions, located 
        farther from the star, the S/N becomes between two and three times higher compared to 
        full-frame PCA. The annular version of PCA does not provide much improvement 
        over full-frame PCA in the small inner working angle region, even for an ADI
        sequence that has a large parallactic angle rotation range ($\sim$80$\degree$) 
        and after adjusting the number of PCs. In this single simulation, we 
        show some practical disadvantages of a full-frame PCA and the gain in S/N we 
        obtain by using local versions of PCA and the proposed three-term decomposition.

\subsection{Performance}        
                        
    \begin{figure*}[!t]
    \begin{center}
    \includegraphics[width=14cm]{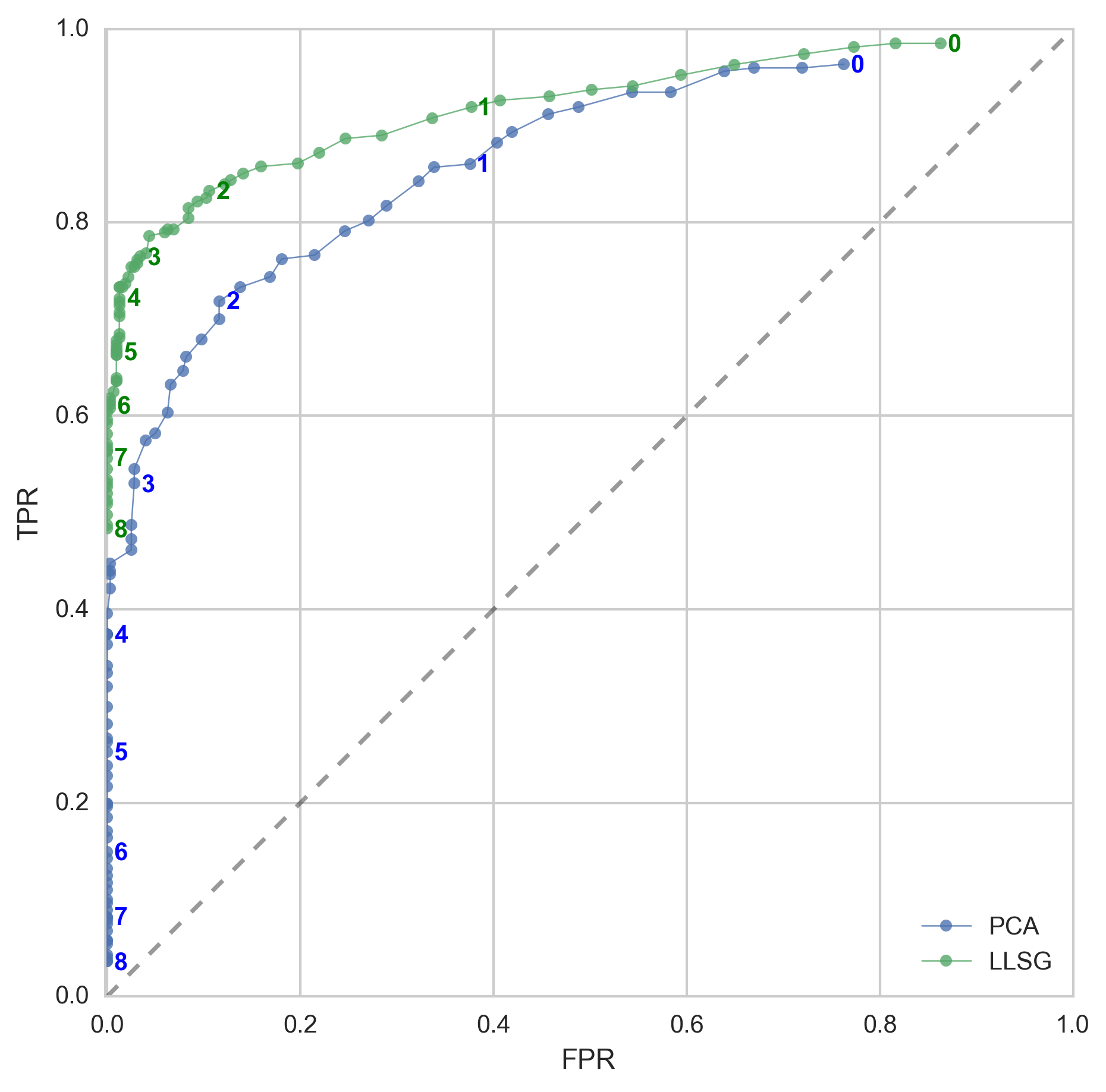}
       \caption{ROC curves for our LLSG decomposition and full-frame PCA.
                    The S/N thresholds $\tau$ are shown for integer values. Our
                    algorithm ROC curve is close to the oracle (perfect classifier)
                    in the upper left corner.}
          \label{fig5}
    \end{center}
    \end{figure*}                               
                        
        Of course, based on a single realization, we cannot characterize the 
        detection performance of the algorithms. More exhaustive approaches are needed, such as the use of receiver operating characteristic (ROC) curves. 
        The performance of a detection algorithm is quantified using ROC analysis, and several meaningful figures of merit can be derived 
        from it \citep{barret06}. 
        ROC and localization ROC curves 
        are widely used tools in statistics and machine learning for visualizing the 
        performance of a binary classifier system in a true positive rate (TPR) - 
        false positive rate (FPR) plot as a decision threshold $\tau$ varies. 
        The ultimate goal of high-contrast imaging,
        as for any signal detection application, is to maximize the TPR while minimizing
        the FPR, which can be achieved by maximizing the area under the curve
        (AUC) in the ROC space \citep{mawet14ss}. In general the goal of a classifier 
        in the ROC space is to be as close as possible to the upper lefthand corner (perfect 
        classifier, referred to as "oracle" with FPR=0 and TPR=1) and away from the random 
        classifier line (TPR=FPR). 
        
        Building this plot for any exoplanet detection algorithm requires a large 
        number of fake companion simulations, especially if a single planet is injected 
        for each realization as in our case. One hundred realizations were made per 
        annulus, centered at 2, 4, 6, 8, 10, and 14 $\lambda/D$, in a random way, meaning 
        that there is a 50$\%$ chance that the datacube contains a synthetic planet. 
        In the case of an injection, the fake companion has a random azimuthal
        position in the given annulus and a random brightness, scaling the PSF (ranging 
        from 0.5 times to 5 times the noise in the considered annulus) as described 
        previously. We consider localization ROC curves for which the decision of 
        whether there is a planet or not is tied to a given position in the image plane.
        
        The detection decision is based on comparing the value of the peak S/N of a given 
        resolution element with a threshold $\tau$. We call the peak S/N here the maximum 
        S/N value obtained from shifting the center of the test resolution element inside
        a $\lambda/D$ circular aperture centered on the considered coordinates. This is
        equivalent to taking the maximum S/N value in a $\lambda/D$ circular aperture, 
        centered on the considered coordinates, from an S/N map. We find this is in practice
        better than using the S/N of a resolution element centered on some given 
        injection coordinates, because the maximum S/N for a point-like source (blob) 
        will usually be shifted by a small amount because of post-processing.  
        A true positive (TP) means that, in the case of an injection, the tested 
        resolution element has a peak S/N $\ge \tau$. A false positive (FP) arises in case
        of a non-injection, when a random resolution element inside the considered 
        annulus has a peak S/N $\ge \tau$. It is important to notice that we inspect
        only one resolution element each time instead of the total number of resolution
        elements in the image (even for the FP count) to preserve the 50-50 prior we 
        described previously. We vary $\tau$ from 0 to 8 in steps of 0.1 in order to 
        have enough points in our empirical ROC curve.
        
        The TPR and FPR for these ROC curves are the averaged TPR and FPR over all 
        brightnesses and the tested annuli. The ROC curves are shown in 
        Fig. \ref{fig5}. It is important to emphasize
        that every point, for every $\tau$, of the LLSG decomposition ROC curve 
        is higher than the one for PCA, which means that the LLSG detection algorithm 
        is closer to the perfect classifier. The full range of values 
        of FPR (up to one) in our ROC curves is not fully covered even when testing 
        unrealistically low values of $\tau$. In this case calculating the AUC becomes
        problematic, and using other metrics derived from a ROC curve becomes more 
        suitable for comparing algorithms (classifiers). An example of such a metric is 
        the Euclidean distance to the upper lefthand corner or "oracle" \citep{braham14}. 
        In the case of PCA, the minimum distance to the upper lefthand corner is 0.3, while for
        our algorithm it is 0.2, which again confirms its superiority.
        
        For generating these ROC curves, we used fifteen PCs, which
        corresponds to 90\% of the explained variance of $M$, a common approach
        for choosing the number of PCs for PCA in machine learning and statistics. 
        The rank of the three-term decomposition was set to fifteen, and the 
        number of iterations was set to ten. Tuning parameters instead of having them 
        fixed for all the realizations could lead to minor improvements in the ROC curves. 
        Tuning the parameters would also increase the complexity in the procedure of 
        generating the ROC curves and would in general be a less fair approach. 
        
\begin{figure}[!t]
    \begin{center}
    \includegraphics[width=9cm]{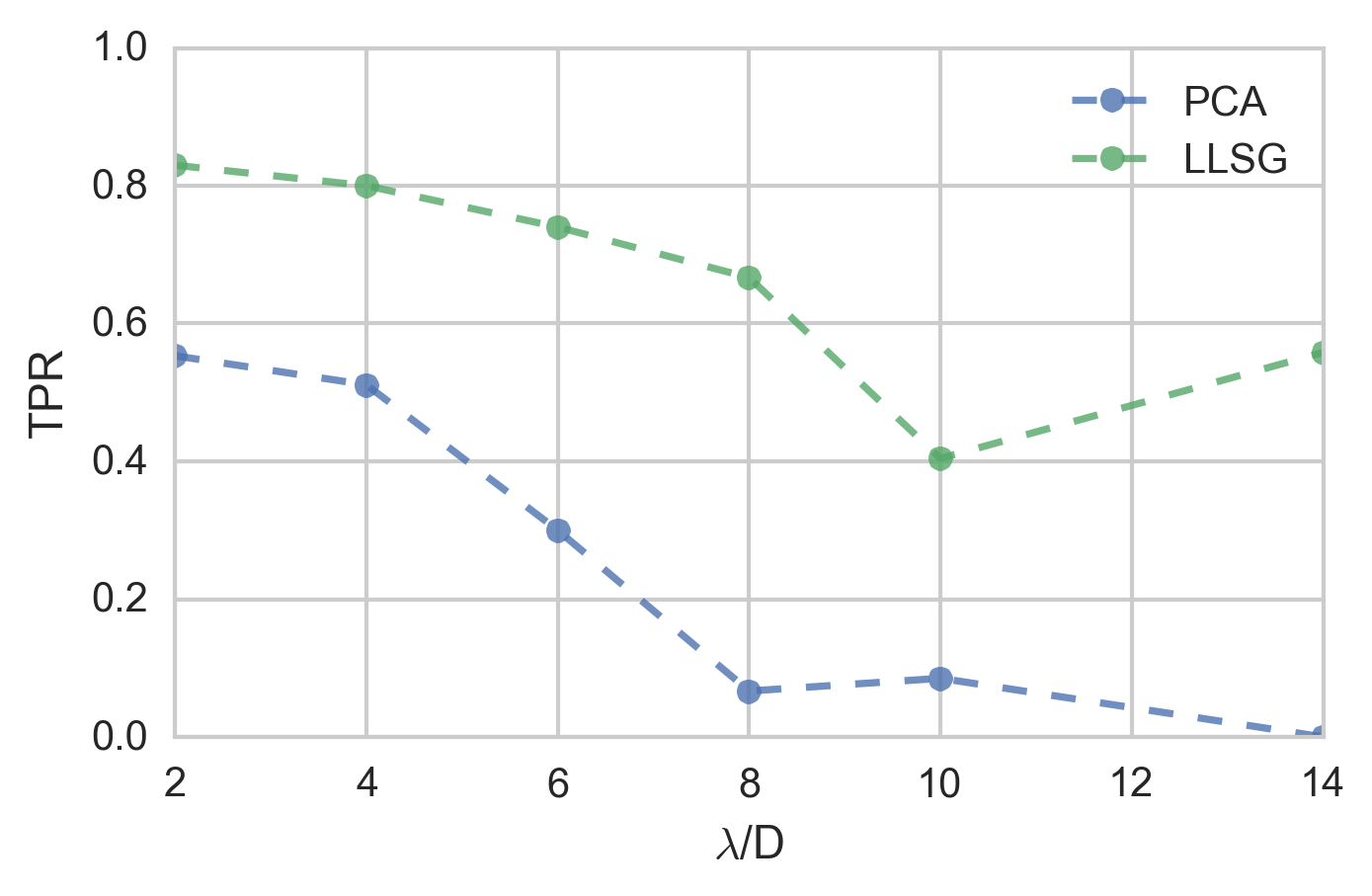}
       \caption{TPR as a function of the distance from the star for an S/N threshold
                        $\tau=5$.}
          \label{fig6}
    \end{center}
    \end{figure}

        The TPR (completeness) is generally a more relevant measure than the FPR, especially 
        for surveys and for obtaining planet population constraints \citep{mawet14ss}. 
        Therefore it is important to evaluate the TPR as a function of the distance 
        from the star for a S/N threshold of 5, which is equivalent to 5-$\sigma$ under 
        the assumption of nearly Gaussian residuals in the final images. The TPR for 
        both algorithms for the tested annuli and $\tau=5$ are shown in Fig. \ref{fig6}. 
        The TPR for the LLSG decomposition is higher for each one of the tested 
        annuli compared to PCA. It is especially interesting how at 2$\lambda/D$, where 
        the speckle noise is dominant, the TPR for our algorithm reaches 83\% instead 
        of the 55\% achieved by PCA. 
                
        Another great advantage of the LLSG decomposition 
        over more expensive algorithms is that its computational cost is comparable to 
        that of full-frame PCA. For instance, it can process the         
        $612\times161\times161$ ($\sim 15.8 \times 10^{6}$) pixel datacube used 
        in our simulations in about ten seconds (when using only one process), whereas 
        full-frame PCA (equivalent to KLIP or pynpoint implementations) using the LAPACK 
        optimized multithreading library can do it in four seconds. 
        This timing depends, as explained before, on the number of iterations for 
        the three-term decomposition.  
        
        It is important to clarify that LLSG is an algorithm for improving detection of
        faint exoplanets, which decomposes the images, separating the static and quasi-static
        structures from the moving planets. This process penalizes the signal of the potential 
        companions, and in consequence the final LLSG frames cannot be used for estimating
        in a robust way the position or flux of those companions. We still need to rely on 
        the injection of negative companion candidates, as we described in a previous section, 
        to calibrate the photometry and astrometry of potential detections, as well as their
        uncertainties.

        In the case of ADI data, the range of rotation (parallactic angles) affects
        the efficiency of post-processing algorithms when searching for potential companions.
        With small rotation, the signal of a planet remains more static through the 
        sequence of frames (this effect gets worse in the innermost part of the frames), 
        and a low-rank approximation based algorithm will fail to retrieve
        it. This effect combines with other factors, such as the number of frames and 
        the PSF decorrelation rate during the sequence, and will limit different 
        post-processing algorithms in different ways. Better understanding of the 
        correlation between these various factors will be useful for choosing 
        algorithms and for designing optimal observing runs.


\section{Conclusions}
        
    In this paper we have shown, for the first time, how recent subspace projection 
    techniques and robust subspace models proposed in the computer vision 
    literature can be applied to ADI high-contrast image sequences. 
        In particular our implementation of a randomized low rank-approximation 
        recently proposed in the machine learning literature coupled with entry-wise 
        thresholding allowed us to decompose an ADI image sequence locally 
    into low-rank, sparse, and noise components. LLSG brings a detectability 
    boost compared to full-frame PCA approach
    at all positions of the field of view as can be seen in the ROC curves with
    averaged TPR and FPR, and in the plot of the TPR as a function of distance.
    This is especially important because it allows us to access the small inner
    working angle region ($\sim 2\lambda/D$ for this dataset), where complex speckle 
    noise prevents PCA from finding faint companions.
        
    One important advantage of this algorithm is that it can process a typical
    $612\times161\times161$ pixel cube without sacrificing too much of the 
    computational cost compared to the fast full-frame PCA approach. 
    That the patches can be processed separately leads to real-time 
    processing if coupled with parallelism to exploit modern multicore 
    architectures, making this algorithm suitable for coming survey pipelines.
    
    We have shown the enormous potential of low-rank plus sparse 
    decompositions and, in particular, the LLSG decomposition for high-contrast 
    imaging. More expensive 
    formulations of these decompositions coupled with a fine-tuned model of the 
    noise could lead to even better reference PSF subtraction for exoplanet 
    detection than the one we proposed in the present paper and will be 
    the focus of future work.

\begin{acknowledgements}
        We would like to thank the anonymous referee for the very useful comments 
        that helped us improve the quality of this paper.
        The research leading to these results has received funding from the European 
    Research Council Under the European Union’s Seventh Framework Program 
    (ERC Grant Agreement n. 337569) and from the French Community of Belgium 
    through an ARC grant for Concerted Research Action.   
\end{acknowledgements}

\bibliographystyle{aa}   
\bibliography{biblio}    
\end{document}